\documentclass[conference]{IEEEtran}
\usepackage[boxruled]{algorithm2e}
\usepackage{cite}
\usepackage{graphicx}
\usepackage{psfrag}
\usepackage{subfigure}
\usepackage{url}
\usepackage{amsmath}
\usepackage{array}
\usepackage{amssymb}
\usepackage{amsfonts}
\usepackage{graphicx}
\usepackage{epstopdf}

\newtheorem{lemma}{Lemma}

\newtheorem{corollary}{Corollary}

\newtheorem{definition}{Definition}
\newtheorem{theorem}{Theorem}
\newtheorem{example}{Example}
\newtheorem{note}{Note}
\title{A Non-Orthogonal DF Scheme for the Single Relay Channel and the Effect of Labelling}
\begin{document}

\author{
\authorblockN{Vijayvaradharaj T Muralidharan}
\authorblockA{Dept. of ECE, Indian Institute of Science \\
Bangalore 560012, India\\
Email: tmvijay@ece.iisc.ernet.in
}
\and
\authorblockN{B. Sundar Rajan}
\authorblockA{Dept. of ECE, Indian Institute of Science, \\Bangalore 560012, India\\
Email: bsrajan@ece.iisc.ernet.in
}
}

\maketitle
\thispagestyle{empty}	
\begin{abstract}
We consider the uncoded transmission over the half-duplex single relay channel, with a single antenna at the source, relay and destination nodes, in a Rayleigh fading environment. The phase during which the relay is in reception mode is referred to as Phase 1 and the phase during which the relay is in transmission mode is referred to as Phase 2. The following two cases are considered: the Non-Orthogonal Decode and Forward (NODF) scheme, in which both the source and the relay transmit during Phase 2 and the Orthogonal Decode and Forward (ODF) scheme, in which the relay alone transmits during Phase 2. A near ML decoder which gives full diversity (diversity order 2) for the NODF scheme is proposed. Due to the proximity of the relay to the destination, the Source-Destination link, in general, is expected to be much weaker than the Relay-Destination link. Hence it is not clear whether the transmission made by the source during Phase 2 in the NODF scheme, provides any performance improvement over the ODF scheme or not. In this regard, it is shown that the NODF scheme provides significant performance improvement over the ODF scheme. In fact, at high SNR, the performance of the NODF scheme with the non-ideal Source-Relay link, is same as that of the NODF scheme with an ideal Source-Relay link. In other words, to study the high SNR performance of the NODF scheme, one can assume that the Source-Relay link is ideal, whereas the same is not true for the ODF scheme. Further, it is shown that proper choice of the mapping of the bits on to the signal points at the source and the relay, provides a significant improvement in performance, for both the NODF and the ODF schemes. 
\end{abstract}
\section{INTRODUCTION}
\label{sec1}
\begin{figure}[htbp]
\centering
\includegraphics[totalheight=.75in,width=1.5in]{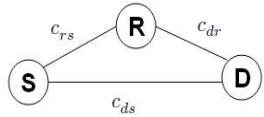}
\caption{The Relay Channel}	
\label{fig:Relay channel}	
\end{figure}
We consider the Rayleigh fading relay channel shown in Fig. \ref{fig:Relay channel}, consisting of the source node $S$, the relay node $R$ and the destination node $D$. It is assumed that R can operate only in the half-duplex mode, i.e., it cannot receive and transmit simultaneously. It is assumed that R has perfect knowledge about the instantaneous value of the fade coefficient associated with the S-R link and D has perfect knowledge about the instantaneous values of the fade coefficients associated with the S-R, R-D and S-D links. Throughout, the phase during which the relay is in reception mode is referred to as Phase 1 and the phase during which the relay is in transmission mode is referred to as Phase 2. In the Non-Orthogonal Decode and Forward (NODF) scheme, S transmits, R and D receive during Phase 1 (Fig. \ref{fig:NODF_Phase1}). Both S and R transmit during Phase 2 (Fig. \ref{fig:NODF_Phase2}). In the Orthogonal Decode and Forward (NODF) scheme, S transmits, R and D receive during Phase 1 (Fig. \ref{fig:ODF_Phase1}). Only R transmits during Phase 2 (Fig. \ref{fig:ODF_Phase2}).

Different decoder architectures for the ODF scheme have been proposed in \cite{SeErAa},\cite{ChLa},\cite{WaCaGiLa} and \cite{JinJinNoShin}. As noted in \cite{SeErAa},\cite{WaCaGiLa} the implementation as well as the performance analysis of the optimal Maximal Likelihood (ML) decoder for the ODF scheme is very complicated. Sub-optimal decoders called $\lambda$-MRC and Co-operative MRC (C-MRC) were proposed in \cite{SeErAa} and \cite{WaCaGiLa} respectively. A near ML decoder for the ODF scheme was presented in \cite{JinJinNoShin} for the single relay channel with multiple antennas. Non-orthogonal relay protocols offer higher spectral efficiency when compared with Orthogonal relay protocols \cite{NaBoKn}, \cite{AzGaSc}, \cite{ElViAnKu}. Power allocation strategies for the NODF scheme were discussed in \cite{YoYoYu}.

In this paper, the near ML decoder presented in \cite{JinJinNoShin} is extended for the NODF scheme. 
The performance of the extended near ML decoder for the NODF scheme is analyzed. Throughout, we consider uncoded communication using signal sets such as M-PSK, M-QAM etc. By a \textit{labelling scheme}, we refer to the way in which the bits are mapped on to the signal points at the source and the relay. Labelling schemes at the source and the relay which result in significant performance improvement are obtained. 

Let $\mathcal{S}=\lbrace s_1,s_1,....,s_M\rbrace$ denote the complex signal set used at S and R, with $\mid\mathcal{S}\mid=M$. A collection of $log_2M$ bits constitutes a message. Let $\mathcal{M}=\lbrace1,2,...,M\rbrace$ denote this message set.
\subsection{Non-Orthogonal Decode and Forward (NODF)}
Let $X_{s_1}:\mathcal{M}\longrightarrow\mathcal{S}$ denote the labelling scheme used at S during Phase 1, i.e., it specifies how messages are mapped onto complex symbols from the signal set at the source. 

\begin{figure}[htbp]
\centering
\includegraphics[totalheight=.75in,width=1.5in]{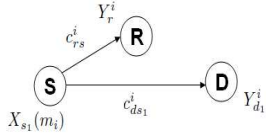}
\caption{The NODF Scheme - Phase 1}	
\label{fig:NODF_Phase1}	
\end{figure}

We assume that  during Phase 1 (Fig. \ref{fig:NODF_Phase1}), S transmits L complex symbols $\lbrace X_{s_1}(m_i)\rbrace_{i=1}^L$, corresponding to L messages $\lbrace m_i\rbrace_{i=1}^L$, where $X_{s_1}(m_i)\in\mathcal{S}$ and $m_i\in\Delta$, for $1 \leq i\leq L$. 

The received signal at R and D during Phase 1 are given by,
\begin {align}
\nonumber
Y_{r}^i&= c_{rs}^i X_{s_1}(m_i)+z_{r}^i,\\
\nonumber
Y_{d_1}^i&= c_{ds_1}^i X_{s_1}(m_i)+z_{d_1}^i,
\end {align}
where $c_{rs}^i$ and $c_{ds_1}^i$ are the zero mean circularly symmetric complex Gaussian fading coefficients associated with the S-R and S-D links respectively with the corresponding variances given by $\sigma_{rs}^2$ and $\sigma_{ds}^2$. The additive noises at R and D, $z_{r}^i$ and $z_{d_1}^i$ are circularly symmetric complex Gaussian random variables with mean 0 and variance 1/2 per dimension, denoted by $CN(0,1)$.

\begin{figure}[htbp]
\centering
\includegraphics[totalheight=.75in,width=1.5in]{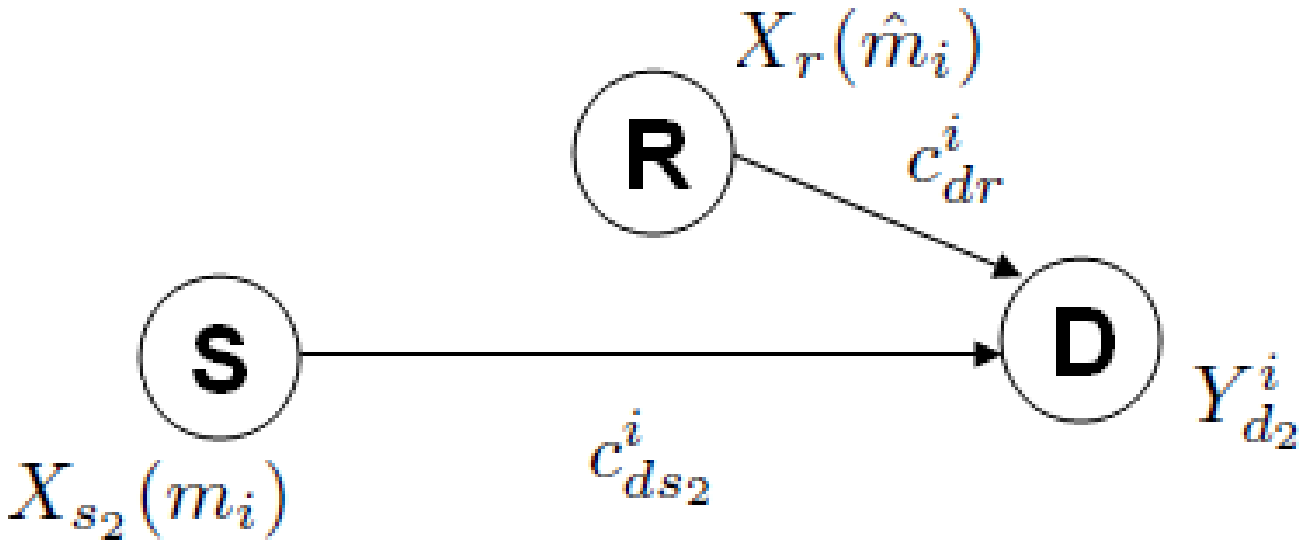}
\caption{The NODF Scheme - Phase 2}	
\label{fig:NODF_Phase2}	
\end{figure}

\begin{figure}[htbp]
\centering
\includegraphics[totalheight=.75in,width=1.5in]{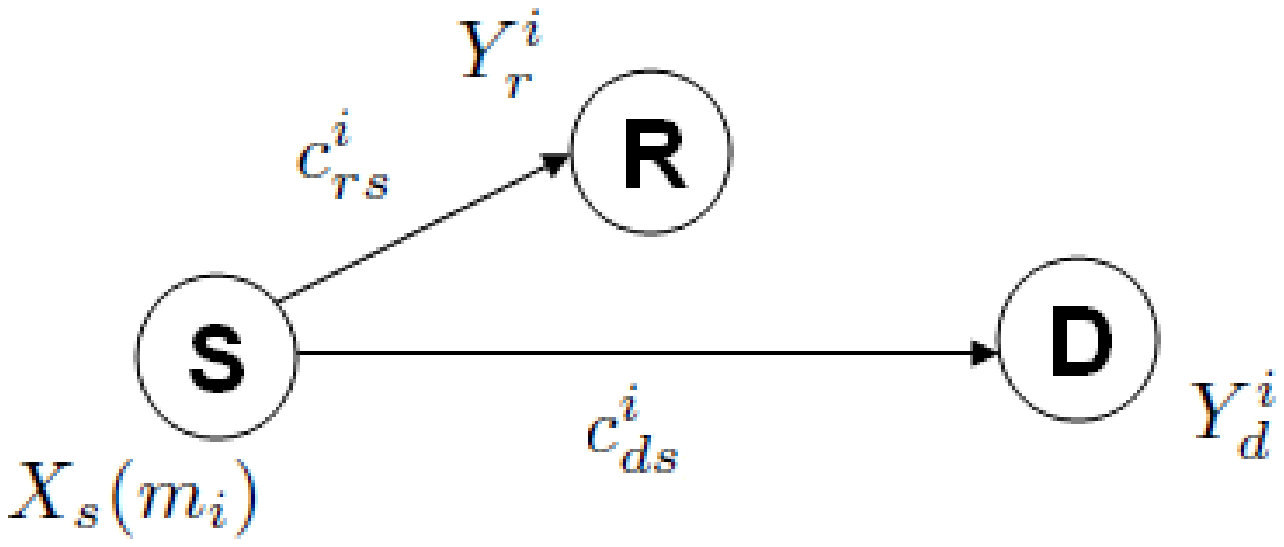}
\caption{The ODF Scheme - Phase 1}	
\label{fig:ODF_Phase1}	
\end{figure}

\begin{figure}[htbp]
\centering
\includegraphics[totalheight=.75in,width=1.5in]{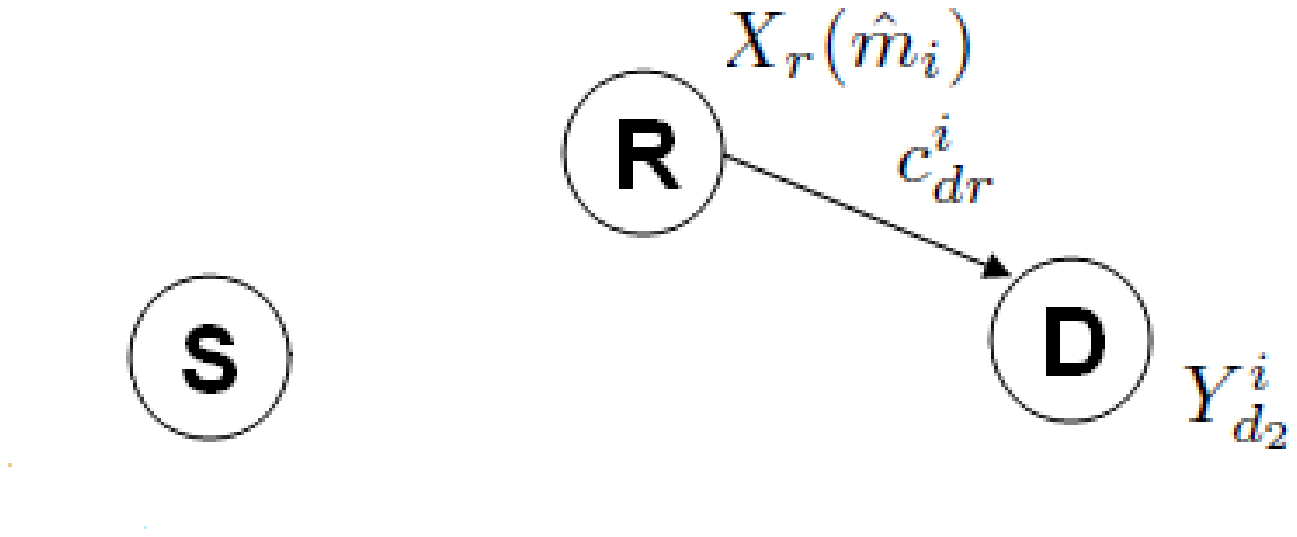}
\caption{The ODF Scheme - Phase 2}	
\label{fig:ODF_Phase2}	
\end{figure}

Let $X_{s_2}:\mathcal{M}\longrightarrow\mathcal{S}$ and $X_{r}:\mathcal{M}\longrightarrow\mathcal{S}$ denote the labelling schemes used at S and R respectively during Phase 2. During Phase 2 (Fig. \ref{fig:NODF_Phase2}), S transmits the L complex symbols $\lbrace X_{s_2}(m_i)\rbrace_{i=1}^L$, corresponding to the same messages $\lbrace m_i\rbrace_{i=1}^L$ transmitted during Phase 1 and R transmits the complex symbols $\lbrace X_{r}(\hat{m}_i)\rbrace_{i=1}^L$ corresponding to the decoded messages $\lbrace \hat{m}_i\rbrace_{i=1}^L$.

The received signal at D during Phase 2 is given by,
\begin {align}
\nonumber
Y_{d_2}^i&= c_{ds_2}^i X_{s_2}(m_i)+c_{dr}^i X_{r}(\hat{m}_i)+z_{d_2}^i,
\end {align}
where $c_{ds_2}^i$ and $c_{dr}^i$ are the zero mean circularly symmetric complex Gaussian fading coefficients associated with the S-D and R-D links respectively with the corresponding variances given by $\sigma_{ds}^2$ and $\sigma_{dr}^2$. The additive noise at D, $z_{d_2}^i$ is $CN(0,1)$.

L is assumed to be large enough such that the fading coefficient associated with the S-D link during Phase 2 $c_{ds_2}^i$ is independent of $c_{ds_1}^i$.
\subsection{Orthogonal Decode and Forward (ODF)}
Let $X_{s}:\mathcal{M}\longrightarrow\mathcal{S}$ denote the labelling scheme used at S during Phase 1.

During Phase 1 (Fig. \ref{fig:ODF_Phase1}), S transmits L complex symbols $\lbrace X_s(m_i)\rbrace_{i=1}^L$, corresponding to L messages $\lbrace m_i\rbrace_{i=1}^L$, where $X_{s}(m_i)\in\mathcal{S}$ and $m_i\in\Delta$, for $1 \leq i\leq L$.
The received signal at R and D during Phase 1 are given by,
\begin {align}
\nonumber
Y_{r}^i&= c_{rs}^i X_{s}(m_i)+z_{r}^i,\\
\nonumber
Y_{d_1}^i&= c_{ds}^i X_{s}(m_i)+z_{d_1}^i,
\end {align}
where $c_{rs}^i$ and $c_{ds}^i$ are the zero mean circularly symmetric complex Gaussian fading coefficients associated with the S-R and S-D links respectively with the corresponding variances given by $\sigma_{rs}^2$ and $\sigma_{ds}^2$. The additive noises at R and D, $z_{r}^i$ and $z_{d_1}^i$ are $CN(0,1)$.

Let $X_{r}:\mathcal{M}\longrightarrow\mathcal{S}$ denote the labelling schemes used at S and R respectively during Phase 2. During Phase 2 (Fig. \ref{fig:ODF_Phase2}), R transmits the complex symbols $\lbrace X_{r}(\hat{m}_i)\rbrace_{i=1}^L$ corresponding to the decoded messages $\lbrace \hat{m}_i\rbrace_{i=1}^L$.

The received signal at D during Phase 2 is given by,
\begin {align}
\nonumber
Y_{d_2}^i&= c_{dr}^i X_{r}(\hat{m}_i)+z_{d_2}^i,
\end {align}
where $c_{dr}^i$ is the zero mean circularly symmetric complex Gaussian fading coefficients associated with the R-D link with the corresponding variance given by $\sigma_{dr}^2$. The additive noise at D, $z_{d_2}^i$ is $CN(0,1)$.

The contributions of the paper are as follows.
\begin{itemize}
\item The expressions for the Pairwise Error Probability (PEP), for the near ML decoder proposed, are derived for the NODF scheme. It is shown that the near ML decoder offers full diversity for the NODF scheme.
\item Even though the S-D link is in general much weaker than the R-D link, the error performance of the NODF scheme is much better than that of the ODF scheme.
\item It is shown that the high SNR performance of the NODF scheme with a non-ideal S-R link, is exactly same as that of the NODF scheme in which the S-R link is ideal. In other words, the effect of the strength of the Source-Relay link completely vanishes at high SNR for the NODF scheme.
\item It is shown that proper choice of the different labelling schemes for the source and the relay, results in a significant improvement in performance over the case where the source and the relay use identical labelling schemes. 
\item Furthermore, it is shown that the performance improvement obtained by proper choice of the labelling scheme is more pronounced in the case of the NODF scheme than the ODF scheme. 
\item  We give an Algorithm to obtain good labelling schemes for the source and the relay.
\end{itemize}

The organization of the rest of the paper is as follows. The description of the near ML decoder for the ODF and NODF schemes constitutes Section II. In Section III, the PEP expressions for the NODF and ODF schemes are derived. In Section IV, the effect of the choice of the labelling scheme on the performance is studied. Section V compares the NODF and ODF schemes with non-ideal Source-Relay link with the case where the Source-Relay link is ideal. In Sections III, IV and V, conclusions derived based on the PEP expressions are validated by simulations, with 8-PSK as the signal set used at the source and the relay. 

\textit{\textbf{Notations:}} 
$CN(0,I_n)$ denotes the standard circularly symmetric complex Gaussian random vector of length $n$. $N(0,c)$ denotes the scalar real valued Gaussian random variable with mean zero and variance $c$. For simplicity, distinction is not made between the random variable and a particular realization of the random variable, in expressions involving probabilities of random variables. For example, $Pr(X=x)$ is simply written as $Pr(X)$. $P_Y(X)$ is the shorthand notation for $Pr(X|Y)$. In some probability expressions involving conditioning of the fading coefficients, the fact that the probability is conditioned on the values taken by the fading coefficients is not explicitly written, as it can be understood from the context. For a set $\mathcal{A}$, $\vert\mathcal{A}\vert$ denotes the cardinality of $\mathcal{A}$. $\Re(x)$ denotes the real part of the complex number $x$. Throughout, $E_{\mathcal{S}}$ denotes the average energy in dB of the signal set $\mathcal{S}$ used at the source and the relay.

\section{A NEAR ML DECODER}

By assumption, R has perfect knowledge about the instantaneous value of the fade coefficient associated with the S-R link and D has perfect knowledge about the instantaneous values of the fade coefficients associated with the S-R, R-D and S-D links.
At R, it is assumed that the decoder performs ML decoding, i.e., the output of the decoder at R
\begin{align}
\nonumber
 \hat{m}_i=\arg \min _{m_i} \vert Y_r^i-c_{rs}^i X_{s_1}(m_i)\vert ^2,
 \end{align} 
for $1 \leq i\leq L$.

For ML decoding at D,  the decoder has to maximize the probability,
\begin{align}
\label{eqn1}
Pr(Y_{d_1}^i,Y_{d_2}^i \vert m_i=a,c_{rs}^i,c_{ds_1}^i,c_{ds_2}^i,c_{dr}^i), 
\end{align}
for $1 \leq i\leq L$, over all possible choices of $a \in \mathcal{M}$.
The form of (\ref{eqn1}) is the same for all $i$. Hence we leave out $i$ in the following discussion. 
The ML decoder decides in favour of $\hat a$, if
\begin{align}
\nonumber
\hat a=\arg \max _a Pr(Y_{d_1},Y_{d_2} \vert m=a,c_{rs},c_{ds_1},c_{ds_2},c_{dr}).
\end{align}
Let $P_a(Y_{d_1},Y_{d_2})=Pr(Y_{d_1},Y_{d_2} \vert m=a,c_{rs},c_{ds_1},c_{ds_2},c_{dr})$.
Then we have,
\begin{align}
\label{eqn2}
P_a(Y_{d_1},Y_{d_2})&=\sum_{j=1}^{M}P_a\left(Y_{d_1},Y_{d_2} \vert X_r\left(j\right)\right)P_a\left(X_r\left(j\right)\right),
\end{align}
where $P_a\left(X_r\left(j\right)\right)$ equals the probability of the event that R decides in favour of message $j \in \mathcal{M}$, given that $a \in \mathcal{M}$ was the message transmitted by the source.
As in \cite{JinJinNoShin}, we upper bound the probability that a message transmitted by S is decoded as another message by the corresponding PEP. Hence, for $j \neq a$, $P_a\left(X_r\left(j\right)\right)$ is upper-bounded by the PEP,
\begin{align}
\nonumber
P_a\left(X_r\left(j\right)\right) &\leq Q\left[\dfrac{\vert c_{rs}\left(X_{s_1}\left(a\right)-X_{s_1}\left(j\right)\right)\vert}{\sqrt{2}}\right]\\
\label{eqn3}
& \leq \dfrac{1}{2}\exp \left\lbrace -\dfrac{1}{4}\left\vert c_{rs}\left(X_{s_1}\left(a\right)-X_{s_1}\left(j\right)\right)\right\vert ^2\right\rbrace,
\end{align}
where Q[.] denotes the complementary CDF of the standard Gaussian random variable.
We also have,
{
\begin{align}
\label{eqn4}
P_a\left(X_r\left(a\right)\right) &\leq 1.
\end{align}
}
For the NODF scheme, the probability $P_a\left(Y_{d_1},Y_{d_2} \vert X_r\left(j\right)\right)$ is given by,
{
\begin{align}
\nonumber
&P_a\left(Y_{d_1},Y_{d_2} \vert X_r\left(j\right)\right)=
\dfrac{1}{\pi} \exp \left \lbrace -\left \vert Y_{d_1}-c_{ds_1} X_{s_1}(a)\right \vert ^2\hspace{30 cm}\right\rbrace\\
 \label{eqn5}
&\hspace{-15 cm}\left\lbrace\hspace{17.8 cm} -\vert Y_{d_2}- c_{ds_2} X_{s_2}(a)-c_{dr}X_{r}({j})\vert ^2 \right \rbrace.
\end{align}
}
Substituting \eqref{eqn3}, \eqref{eqn4} and \eqref{eqn5} in \eqref{eqn2}, we get \eqref{eqn6} (shown at the top of the next page).
As in \cite{JinJinNoShin}, the near ML decoder tries to maximize the dominant exponential in the upper bound \eqref{eqn6}.
\begin{figure*}
\scriptsize
\begin{align}
\nonumber
P_a(Y_{d_1},Y_{d_2})&\leq
\dfrac{1}{\pi}\exp \left \lbrace -\left \vert  Y_{d_1}-c_{ds_1} X_{s_1}(a)\right \vert ^2 
-\vert Y_{d_2}- c_{ds_2} X_{s_2}(a)+c_{dr}X_{r}({a})
\vert ^2 \right \rbrace \\
\label{eqn6}
&+\dfrac{1}{2\pi}\sum_{j=1,j \neq a}^{M} \left(\exp \left \lbrace -\left \vert  Y_{d_1}-c_{ds_1} X_{s_1}(a)\right \vert ^2 -\vert Y_{d_2}- c_{ds_2} X_{s_2}(a)+c_{dr}X_{r}({j})
\vert ^2 \right \rbrace \exp\left\lbrace\ -\dfrac{1}{4}\vert c_{rs}\left(X_{s_1}\left(a\right)-X_{s_1}\left(j\right)\right)\vert^2\right\rbrace\right)
\end{align}
\hrule
\end{figure*}

Let us define,
\begin{align}
\nonumber
f^j(a)&=\dfrac{1}{4}\left\vert c_{rs}\left(X_{s_1}\left(a\right)-X_{s_1}\left(j\right)\right)\right\vert ^2 \\
\nonumber
&+\left \vert  Y_{d_1}-c_{ds_1} X_{s_1}(a)\right \vert ^2 +\vert Y_{d_2}- c_{ds_2} X_{s_2}(a)-c_{dr}X_{r}({j})\vert ^2,
\end{align}
for $1 \leq j \leq M$ and $1 \leq a \leq M$.
The near ML decoder decides in favour of
{
\begin{align}
\nonumber
\hat{a}=\arg \min_{a=1,2,...,M} \left \lbrace \min_{i=1,2,...,M}f^i\left(a\right)\right\rbrace.
\end{align} 
}
In other words, $\hat{a}=\bar{a}$, if
{
\begin{align}
\nonumber
\min_{a=1,2,...,M} \left \lbrace \min_{i=1,2,...,M}f^i\left(a\right)\right\rbrace = f^k(\bar{a}),
\end{align}
}
for some $1 \leq k \leq M$.

The near ML decoder for the ODF scheme can be obtained as a special case of the NODF scheme by taking $c_{ds_1}=c_{ds}$, $X_{s_1}(a)=X_{s}(a)$ and $X_{s_2}(a)=0$, for all $a \in \mathcal{M}$.
\section{PERFORMANCE ANALYSIS OF THE NEAR ML DECODER}
The following Lemma is useful for the performance analysis of the near ML decoder.
\begin{lemma}
\label{lemma1}
Let $x_1$ and $x_2$ $\in$ $\mathbb{C}^n$ be the transmitted vectors corresponding to messages 1 and 2 respectively. The received vector $y=x+z$ $\in$ $\mathbb{C}^n$, where $x$ $\in$ $\lbrace x_1,x_2 \rbrace$  and $z$ is $CN(0,I_n)$. The decoder decides in favour of message $1$,
if $||y-x_1||^2 \leq ||y-x_2||^2 + c$.
Otherwise, it decides in favour of message $2$,
where $c \in \mathbb{R}$ is a constant. Then the probability that the decoder decides in favour of message $2$, given that message $1$ was transmitted is upper bounded as,

\begin{align}
\nonumber
Pr \left(1 \longrightarrow 2 \right) \leq \dfrac{1}{2} \exp \left \lbrace -\dfrac{||x_1-x_2||^2}{4}-\dfrac{c}{2}\right\rbrace.
\end {align}

\begin{proof}

We have,
\begin{align}
\nonumber
Pr \left(1 \longrightarrow 2 \right)&=P_1\left(||y-x_1||^2 \geq ||y-x_2||^2 + c\right)\\
\nonumber
&=P_1\left(\Re\left \lbrace \left (y- \dfrac {x_1+x_2} {2}\right)^* \left(x_2 - x_1)\right)  \right\rbrace \geq \dfrac{c}{2}\right)\\
\nonumber
&=Pr\left(\Re\left \lbrace \left (z+\dfrac {x_1-x_2} {2}\right)^* \left(x_2 - x_1)\right)  \right\rbrace \geq \dfrac{c}{2}\right)\\
\nonumber
&=Pr\left(\Re\left \lbrace z^* \dfrac {x_2-x_1}{\vert\vert x_2 - x_1 \vert\vert} \right\rbrace\hspace{30 cm}\right) \\
\nonumber
&\hspace{-15 cm}\left(\hspace{17.4 cm}\geq \dfrac{c}{2{\vert\vert x_2 - x_1 \vert\vert}}+{\dfrac{\vert\vert x_2 - x_1 \vert\vert}{2}}\right)
\end{align}
Since $z$ is $CN$(0,$I_n$), it can be shown that $\Re\left\lbrace z^*\dfrac {x_2-x_1}{\vert\vert x_2 - x_1 \vert\vert}\right\rbrace$ is  $N\left(0,\dfrac{1}{2}\right)$.
\newline
Hence,
\begin{align}
\nonumber
Pr \left(1 \longrightarrow 2 \right)&=Q\left[\sqrt{2}\left(\dfrac{\vert\vert x_1-x_2 \vert \vert}{2}+\dfrac{c}{2{\vert\vert x_2 - x_1 \vert\vert}}\right)\right]\\
\nonumber
&\leq\exp\left\lbrace-\left(\dfrac{\vert\vert x_1-x_2 \vert \vert}{2}+\dfrac{c}{2{\vert\vert x_2 - x_1 \vert\vert}}\right)^2\right\rbrace\\
\nonumber
&\leq \exp \left \lbrace -\dfrac{||x_1-x_2||^2}{4}-\dfrac{c}{2}\right\rbrace.
\end{align}
\end{proof}
\end{lemma}
\begin{theorem}
For the NODF scheme, the PEP that the decoder at D decides in favour of message $\bar{a} \in \mathcal{M}$ given that the message transmitted by the source was $a \in \mathcal{M}$ is upper bounded as,
\begin{align}
\nonumber
&Pr \left(a \longrightarrow \bar{a} \right) \leq \\
\nonumber
&\hspace{0 cm}\dfrac{1}{2} \left [\dfrac{1}{1+\dfrac{1}{4}\vert \sigma_{ds}\vert ^2\vert X_{s_1}({a})-X_{s_1}(\bar{a})\vert ^2}\right]\\
\nonumber
&\left[\dfrac{1}{1+\dfrac{1}{4}\vert \sigma_{ds}\vert ^2\vert X_{s_2}({a})-X_{s_2}(\bar{a})\vert ^2+\dfrac{1}{4}\vert \sigma_{dr}\vert ^2\vert X_{r}({a})-X_{r}(\bar{a})\vert ^2}\right]\\
\nonumber
&\hspace{7.7 cm} + H.O.T.
\end{align}
where H.O.T denotes the terms of order greater than 2.

\begin{proof}

{
We have,
\begin{align}
\nonumber
Pr \left(a \longrightarrow \bar{a} \right)&=Pr \left(a \longrightarrow \bar{a} \right\vert X_r\left(a\right))P_a\left(X_r\left(a\right)\right)\\
\label{eqn66} 
&+\sum_{j=1,j\neq a}^M Pr \left(a \longrightarrow \bar{a} \right\vert X_r\left(j\right))P_a\left(X_r\left(j\right)\right).
\end{align}
Also,
\begin{align}
\label{eqn7}
P_a\left(X_r\left(j\right)\right) \leq \dfrac{1}{2} \exp\left\lbrace-\dfrac{1}{4}\vert c_{rs}\left(X_s\left(a\right)-X_s\left(j\right)\right)\vert^2 \right\rbrace,
\end{align}
and
\begin{align}
\label{eqn9}
P_a\left(X_r\left(a\right)\right) \leq 1.
\end{align}
Further, we have,
\begin{align}
\label{eqn10}
Pr\left(a \longrightarrow \bar{a} \vert X_r\left(a\right)\right) \leq \sum_{l=1}^{M}P_a\left(f^a\left(a\right) \geq f^l\left(\bar{a}\right)\right).
\end{align}
Using Lemma \ref{lemma1} in \eqref{eqn10} gives rise to \eqref{eqn11} (shown at the top of the next page).
Similarly we have,
\begin{align}
\label{eqn12}
Pr\left(a \longrightarrow \bar{a} \vert X_r\left(j\right)\right) \leq \sum_{m=1}^{M}P_a\left(f^j\left(a\right) \geq f^m\left(\bar{a}\right)\right);
\end{align}
Using Lemma \ref{lemma1} in \eqref{eqn12} gives rise to \eqref{eqn13} (shown at the top of the next page).
\begin{figure*}
\scriptsize
\begin{align}
\label{eqn11}
&Pr\left(a \longrightarrow \bar{a} \vert X_r\left(a\right)\right) \leq \dfrac{1}{2}\sum_{l=1}^{M}\exp\left\lbrace -\dfrac{\vert c_{ds_1}
\left(X_{s_1}\left(a\right)-X_{s_1}\left(\bar{a}\right)\right)\vert^2}{4}
-\dfrac{\vert c_{ds_2}
\left(X_{s_2}\left(a\right)-X_{s_2}\left(\bar{a}\right)\right)+ c_{dr}
\left(X_{r}\left(a\right)-X_{r}\left(l\right)\right)\vert^2}{4}-\dfrac{\vert c_{rs}\left(X_{s_1}\left(\bar{a}\right)-X_{s_1}\left(l\right)\right)\vert^2}{8}\right\rbrace\\
\nonumber
&Pr\left(a \longrightarrow \bar{a} \vert X_r\left(j\right)\right) \leq \dfrac{1}{2}\sum_{m=1}^{M}\exp\left\lbrace -\dfrac{\vert c_{ds_1}
\left(X_{s_1}\left(a\right)-X_{s_1}\left(\bar{a}\right)\right)\vert^2}{4}
-\dfrac{\vert c_{ds_2}
\left(X_{s_2}\left(a\right)-X_{s_2}\left(\bar{a}\right)\right)+ c_{dr}
\left(X_{r}\left(j\right)-X_{r}\left(m\right)\right)\vert^2}{4}\hspace{20 cm}\right\rbrace\\
\label{eqn13}
&\hspace{-15 cm}\left\lbrace \hspace{25 cm}-\dfrac{\vert c_{rs}\left(X_{s_1}\left(\bar{a}\right)-X_{s_1}\left(m\right)\right)\vert^2}{8}
+\dfrac{\vert c_{rs}\left(X_{s_1}\left({a}\right)-X_{s_1}\left(j\right)\right)\vert^2}{8}\right\rbrace\\
\nonumber
&Pr(a \longrightarrow \bar{a}) \leq \dfrac{1}{2} \sum_{l=1}^M\left(\left[\dfrac{1}{1+\dfrac{\vert \sigma_{ds}
\left(X_{s_1}\left(a\right)-X_{s_1}\left(\bar{a}\right)\right)\vert^2}{4}}\right]
\left[\dfrac{1}{1+\dfrac{\vert \sigma_{ds}
\left(X_{s_2}\left(a\right)-X_{s_2}\left(\bar{a}\right)\right)\vert^2}{4} + \dfrac{\vert \sigma_{dr}
\left(X_{r}\left(a\right)-X_{r}\left({l}\right)\right)\vert^2}{4}}\right]\hspace{30 cm}\right)\\
\nonumber
&\hspace{-15 cm}\left(\hspace{27.8 cm}\left[\dfrac{1}{1+\dfrac{\vert \sigma_{rs}
\left(X_{s_1}\left(\bar{a}\right)-X_{s_1}\left(l\right)\right)\vert^2}{8}}\right]\hspace{0 cm}\right)\\
\nonumber
&\hspace{2 cm}+\dfrac{1}{4}\sum_{j=1,j \neq a}^{M}\sum_{m=1}^M\left(\left[\dfrac{1}{1+\dfrac{\vert \sigma_{ds}
\left(X_{s_1}\left(a\right)-X_{s_1}\left(\bar{a}\right)\right)\vert^2}{4}}\right]
\left[\dfrac{1}{1+\dfrac{\vert \sigma_{ds}
\left(X_{s_2}\left(a\right)-X_{s_2}\left(\bar{a}\right)\right)\vert^2}{4} + \dfrac{\vert \sigma_{dr}
\left(X_{r}\left(a\right)-X_{r}\left({m}\right)\right)\vert^2}{4}}\right]\hspace{30 cm}\right)\\
\label{eqn14}
&\hspace{-20 cm}\left(\hspace{29 cm}\left[\dfrac{1}{1+\dfrac{\vert \sigma_{rs}
\left(X_{s_1}\left({a}\right)-X_{s_1}\left(j\right)\right)\vert^2}{8}+\dfrac{\vert \sigma_{rs}
\left(X_{s_1}\left(\bar{a}\right)-X_{s_1}\left(m\right)\right)\vert^2}{8}}\right]\right)
\end{align}
\hrule
\end{figure*}
Substituting \eqref{eqn7}, \eqref{eqn9}, \eqref{eqn11} and \eqref{eqn13} in \eqref{eqn66} and taking expectation with respect to $c_{rs}$, $c_{ds_1}$, $c_{ds_2}$ and $c_{dr}$, we get \eqref{eqn14} (shown at the top of the next page).

Neglecting terms which are of order 3 in \eqref{eqn14} gives the result.

}
\end{proof}
\end{theorem}
\begin{corollary}
For the ODF scheme, the PEP that the decoder at D decides in favour of message $\bar{a} \in \mathcal{M}$ given that the message transmitted by the source was $a \in \mathcal{M}$ is upper bounded as,
\begin{align}
\nonumber
&Pr \left(a \longrightarrow \bar{a} \right) \leq \\
\nonumber
&\hspace{0 cm} \left [\dfrac{1}{1+\frac{1}{4}\vert \sigma_{ds}\vert ^2\vert X_{s}({a})-X_{s}(\bar{a})\vert ^2}\right] 
\left[\dfrac{1}{1+\frac{1}{4}\vert \sigma_{dr}\vert ^2\vert X_{r}({a})-X_{r}(\bar{a})\vert ^2}\right]\\
\nonumber
&\hspace{4.5 cm}+
\\
\nonumber
&\hspace{0 cm} \left [\dfrac{1}{1+\frac{1}{4}\vert \sigma_{ds}\vert ^2\vert X_{s}({a})-X_{s}(\bar{a})\vert ^2}\right]\\
\nonumber 
&\sum_{j=1,j\neq a}^{M}\left[\dfrac{1}{1+\frac{1}{8}\vert \sigma_{rs}\vert ^2\left(\vert X_{s}({a})-X_{s}(j)\vert ^2+\vert  X_{s}({a})-X_{s}(\bar{a})\vert ^2\right)}\right]\\
\nonumber
&\hspace{7.7 cm} + H.O.T.
\end{align}
where H.O.T denotes the terms of order greater than 2.

\begin{proof}

Substituting $X_{s_2}(n)=0$, for all $n \in \mathcal{M}$, and neglecting terms of order 3 in \eqref{eqn14} gives the result.
\end{proof}
\end{corollary}

\begin{note}
From the bounds on the PEP given by Theorem 1 and Corollary 1, it is clear that relative rotation of the signal sets if used at S and R will have no effect on the error performance of the NODF and the ODF schemes.
\end{note}

\begin{figure}[htbp]
\centering
\includegraphics[totalheight=2.5in,width=3.75in]{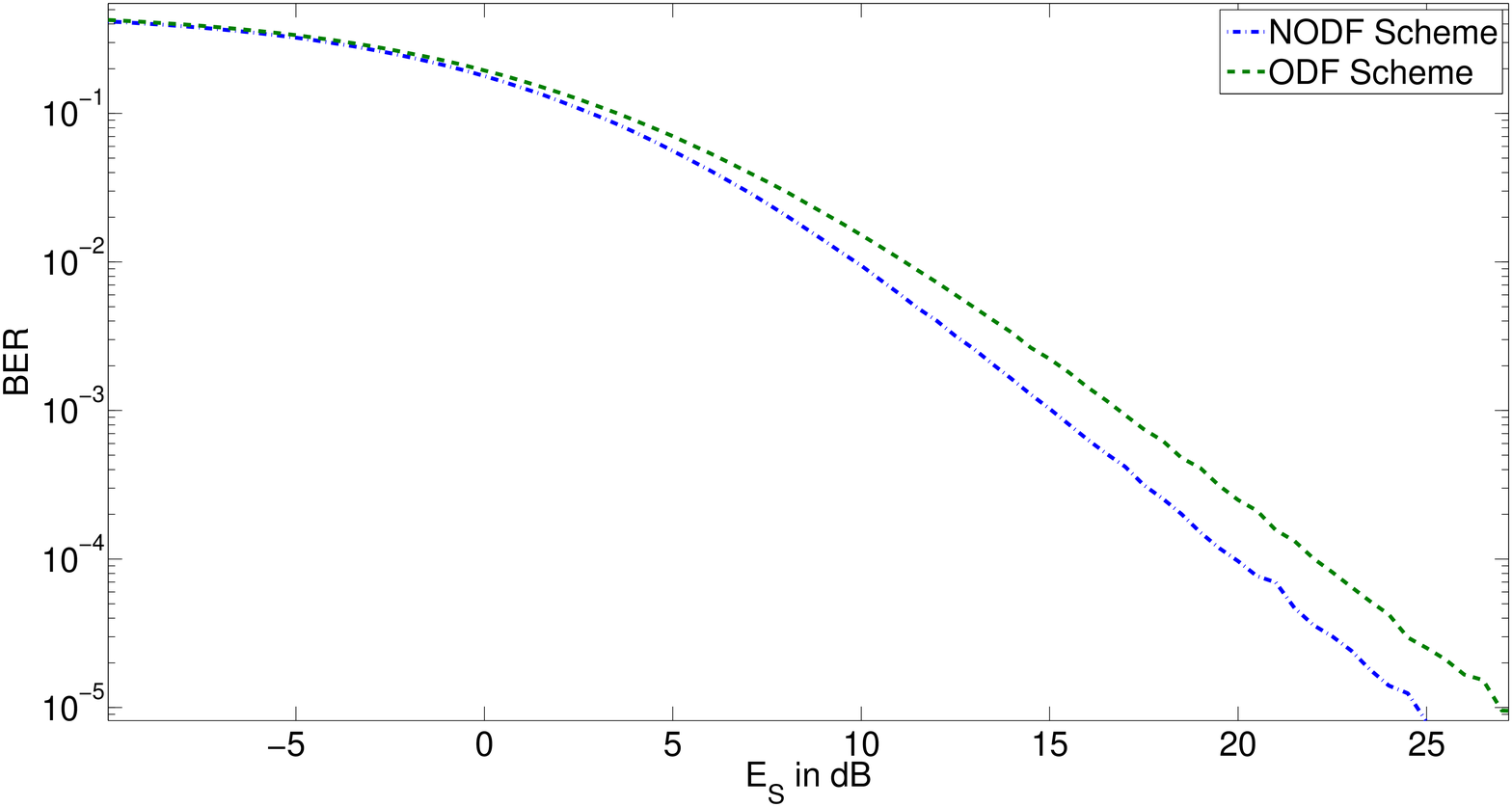}
\caption{$E_{\mathcal{S}}$ Vs $BER$ performance of the NODF and the ODF schemes without our labelling for 8-PSK, with $\sigma_{ds}=0$ dB, $\sigma_{rs}=10$ dB and $\sigma_{dr}=10$ dB.}	
\label{fig:fig1}	
\end{figure}

\begin{figure}[htbp]
\centering
\includegraphics[totalheight=2.5in,width=3.75in]{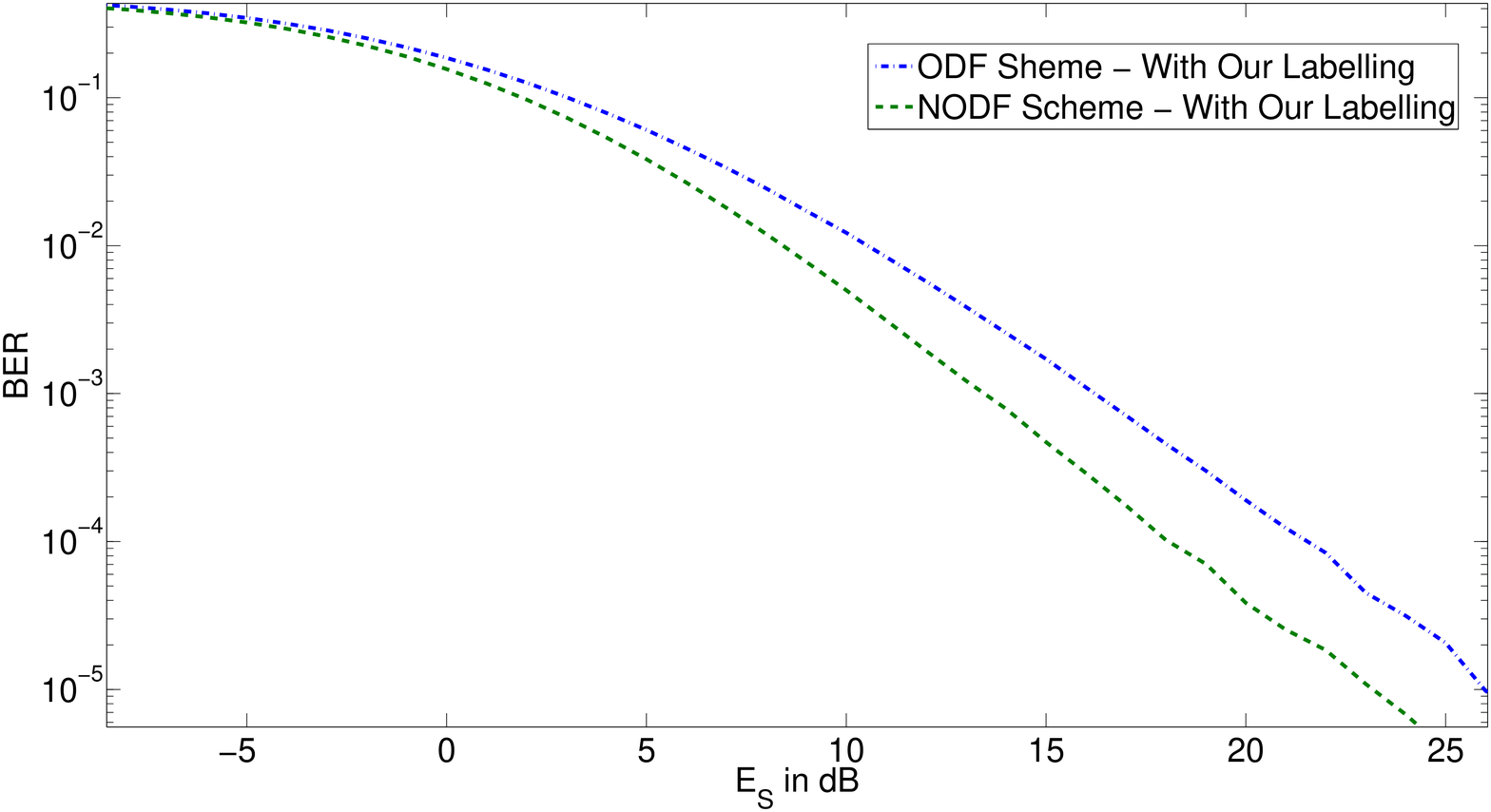}
\caption{$E_{\mathcal{S}}$ Vs $BER$ performance of the NODF and the ODF schemes with our labelling for 8-PSK, with $\sigma_{ds}=0$ dB, $\sigma_{rs}=10$ dB and $\sigma_{dr}=10$ dB.}	
\label{fig:fig2}	
\end{figure}

From Theorem 1 and Corollary 1, it follows that the diversity order of the NODF and ODF schemes, for the near ML decoder proposed is 2. We see that for the NODF scheme, the PEP has only one term of order 2, whereas the PEP of the ODF scheme has additional terms of order 2. This leads to the following conclusion. Even though the S-D link in general is much weaker than the R-D link, the transmission made by the source during Phase 2 in the NODF scheme is indeed beneficial, since it results in a lesser PEP. The simulation results showing the $E_{\mathcal{S}}$ Vs $BER$ performance of the NODF and ODF schemes with 8-PSK as the input constellation at S and R are presented in Fig. \ref{fig:fig1} and Fig. \ref{fig:fig2}. The variances of the fading coefficients are assumed to be $\sigma_{ds}=0$ dB, $\sigma_{rs}=10$ dB and $\sigma_{dr}=10$ dB.
In Fig. \ref{fig:fig1}, the simulation results shown are for the case where the labelling scheme used by S and R are the same. From Fig. \ref{fig:fig1}, we see that at high SNR, using the NODF scheme provides an advantage of 1.5 dB over the ODF scheme. In Fig. \ref{fig:fig2}, the simulation results shown are for the case where S and R use the labelling scheme described in Section V. From Fig. \ref{fig:fig2}, we see that for this case, at high SNR, using the NODF scheme provides an advantage of 3.5 dB over the ODF scheme. 

\section{CHOICE OF THE LABELLING SCHEME}
From Theorem 1, it follows that in order to minimise the PEP for the NODF scheme, we need to maximize the following:
{
\begin{align}
\nonumber
&\hspace{0 cm} \left [{1+\dfrac{1}{4}\vert \sigma_{ds}\vert ^2\vert X_{s_1}({a})-X_{s_1}(\bar{a})\vert ^2}\right]\\
\nonumber
&\left[{1+\dfrac{1}{4}\vert \sigma_{ds}\vert ^2\vert X_{s_2}({a})-X_{s_2}(\bar{a})\vert ^2+\dfrac{1}{4}\vert \sigma_{dr}\vert ^2\vert X_{r}({a})-X_{r}(\bar{a})\vert ^2}\right].
\end{align}
}
At high SNR the metric we need to maximize becomes,
\begin{align}
\nonumber
m(a,\bar{a})=&{\vert X_{s_1}({a})-X_{s_1}(\bar{a})\vert ^2}\\
\nonumber
&\left[{\alpha\vert X_{s_2}({a})-X_{s_2}(\bar{a})\vert ^2+\vert X_{r}({a})-X_{r}(\bar{a})\vert ^2}\right],
\end{align}
where $\alpha=\dfrac{\vert \sigma_{ds}\vert ^2}{\vert \sigma_{dr}\vert ^2}$.
Throughout, we assume $\alpha \ll 1$, since in general $\vert\sigma_{ds}\vert \ll \vert\sigma_{dr}\vert$.
Let $m_1(a,\bar{a})=\vert X_{s_1}({a})-X_{s_1}(\bar{a})\vert ^2$, $m_2(a,\bar{a})=\vert X_{r}({a})-X_{r}(\bar{a})\vert ^2$, $m_3(a,\bar{a})=\vert X_{s_2}({a})-X_{s_2}(\bar{a})\vert ^2$.
In order to minimize the error probability, for the NODF scheme, we need to maximize the minimum value of $m(a,\bar{a})$ over all possible choices of the message pairs $(a,\bar{a})$. For the ODF scheme we note that only the first term in the PEP expression in Corollary 1 involves $X_r$. As a result, the metric for the ODF scheme can be obtained as a special case by choosing $X_{s_2}(n)=0$ for all $n \in \mathcal{M}$. 

For the ODF scheme, from Corollary 1, the PEP expression has $M$ terms of order 2. The first term is dependent on the choice of the labelling scheme and the other terms are independent of the choice of the labelling scheme. Hence the effect of the choice of the labelling scheme is expected to be less pronounced for the ODF scheme when compared with the NODF scheme.   

Let $\mathcal{L}$ denote the labelling scheme used at S and R, and also let
\begin{align}
\nonumber
p(a)=\min_{\bar{a}, \bar{a} \neq a} m(a,\bar{a}). 
\end{align}
Let us define,
\begin{align}
\nonumber
d(\mathcal{L})&=\min_{a,\bar{a}, a \neq \bar{a}} m(a,\bar{a}) = \min_a p(a).
\end{align}

Let $\mathcal{L}_0$ denote the labelling scheme in which the mapping from bits to complex symbols, used by S (during Phase 1 and Phase 2) and R (during Phase 2), are the same. Similar to $p(a)$ and $d(\mathcal{L})$, $p_0(a)$ and $d(\mathcal{L}_0)$ are defined for the labelling scheme $\mathcal{L}_0$.
\begin{definition}
The Labelling Gain of the labelling scheme $\mathcal{L}$, which is a measure of the performance gain provided by $\mathcal{L}$ over $\mathcal{L}_0$, is given by,
\newline
$L_G(\mathcal{L})=10\log_{10}\left[{\dfrac{d(\mathcal{L})}{d(\mathcal{L}_0)}}\right]$ dB.
\end{definition}
It is important to note that the Labelling Gain is calculated based on the upper bound on the PEP, taking into consideration only those pair of messages $a$ and $\bar{a}$ which contribute dominantly to the metric $m(a,\bar{a})$. The actual high SNR gain provided by the labelling scheme  $\mathcal{L}$ over the scheme $\mathcal{L}_0$ need not equal $L_G(\mathcal{L})$.

Throughout, the phrase \textit{with our labelling} means that S and R use the labelling scheme which is to be described in this section and  \textit{without our labelling} means that S and R use the labelling scheme $\mathcal{L}_0$.

An Algorithm to obtain a good labelling scheme is as follows.
\begin{itemize}
\item Choice of $X_{s_1}$ :
\newline
The mapping $X_{s_1}$ used by S during Phase 1 can be chosen arbitrarily. In particular, we can choose the mapping in which message $i$ is mapped on to $s_i$.
\item Choice of $X_{r}$ :
\newline
\underline{Step 1:}
\newline
$X_r(1)$ can be chosen arbitrarily. In particular we can choose $X_r(1)=s_1$.
\newline
\underline{Step 2:}
\newline
By the choice of $X_{s_1}$, $\min_j m_1(i,j)$ occurs for a set of values of $j$, denoted as $\mathcal H_i$.  Assign symbols $X_r(j)$ for $j \in \mathcal H_1$, in the increasing order of $j$, such that $m_2(1,j)$ is maximum, i.e, choose $X_r(j)=s'$, where $s'$ is chosen  to be the one which has the maximum Euclidean distance from $s_1$, among all symbols of $\mathcal{S}$ which are not previously assigned. If more than one option is available while making a choice, choose any one.
\newline
\underline{Step 3}:
\newline
Consider the sets $H_l$, where $l$ belongs the set of messages for which symbols have been assigned. For each one of the sets, assign symbols $X_r(j_l)$ for $j_l \in \mathcal H_l$, such that $X_r(j_l)=s'$, where $s'$ is chosen  to be the one which has the maximum Euclidean distance from $X_r(l)$, among all symbols of $\mathcal{S}$ which are not previously assigned.
\newline
\underline{Step 4}:
\newline
Repeat Step 3 for those messages for which symbols have not been assigned.
\newline
\underline{Step 5}:
\newline
If the procedure described above results in a value of $\min_{a,\bar{a}}m_1(a,\bar{a})m_2(a,\bar{a})=\delta^2$, where $\delta$ is the minimum of the squared Euclidean distance between all pairs of points in the signal set $\mathcal{S}$, change the choice which was made recently and repeat Steps 3 and 4 to ensure that $\min_{a,\bar{a}}m_1(a,\bar{a})m_2(a,\bar{a})$ is greater than $\delta^2$.
\item Choice of $X_{s_2}$
\newline
\underline{Step 1:}
\newline
$X_{s_2}(1)$ can be chosen arbitrarily. In particular we can choose $X_{s_2}(1)=s_1$.
\newline
\underline{Step 2:}
\newline
By the choice of $X_{s_1}$ and $X_{r}$, $\min_j m_1(i,j)m_2(i,j)$ occurs for a set of values of $j$, denoted as $\mathcal K_i$.  Assign symbols $X_{s_2}(j)$ for $j \in \mathcal K_1$, in the increasing order of $j$, such that $m_3(1,j)$ is maximum, i.e, choose $X_{s_2}(j)=s'$, where $s'$ is chosen  to be the one which has the maximum Euclidean distance from $s_1$, among all symbols of $\mathcal{S}$ which are not previously assigned. If more than one option is available while making a choice, choose any one.
\newline
\underline{Step 3}:
\newline
Consider the sets $K_l$, where $l$ belongs the set of messages for which symbols have been already assigned. For each one of the sets, assign symbols $X_{s_2}(j_l)$ for every $j_l \in \mathcal K_l$, such that $X_r(j_l)=s'$, where $s'$ is chosen  to be the one which has the maximum Euclidean distance from $X_{s_2}(l)$, among all symbols of $\mathcal{S}$ which are not previously assigned.
\newline
\underline{Step 4}:
\newline
Repeat Step 3 for those messages for which symbols have not been assigned.
\newline

The Algorithm to find a good labelling strategy for the ODF scheme, i.e., choosing the maps $X_{s}$ and $X_r$ are exactly same as the choice of the maps $X_{s_1}$ and $X_r$ for the NODF scheme.  
\end{itemize}
\begin{table*}
\centering
\caption{4-PSK, NODF SCHEME}
\label{table:table1}
\begin{tabular}{|c|c|c|c|c|c|c|c|c|}
\hline Bits & Message j & $X_{s_1}(j)$ & $\mathcal{H}_j$ & $X_r(j)$ & $\mathcal{K}_j$ & $X_{s_2}(j)$ & $p(j)$ & $p_0(j)$ \\ 
\hline 00 & 1 & $s_1$ & $\lbrace2,4\rbrace$ & $s_1$ & $\lbrace4\rbrace$ & $s_1$ & $4.8$ & $4.4$ \\ 
\hline 01 & 2 & $s_2$ & $\lbrace1,3\rbrace$ & $s_3$ & $\lbrace3\rbrace$ & $s_2$ & $4.8$ & $4.4$ \\ 
\hline 10 & 3 & $s_3$ & $\lbrace2,4\rbrace$ & $s_4$ & $\lbrace2\rbrace$ & $s_4$ & $4.8$ & $4.4$ \\ 
\hline 11 & 4 & $s_4$ & $\lbrace3,1\rbrace$ & $s_2$ & $\lbrace1\rbrace$ & $s_3$ & $4.8$ & $4.4$ \\ 
\hline 
\end{tabular} 
\end{table*}
\begin{table*}
\centering
\caption{4-PSK, ODF SCHEME}
\label{table:table2}
\begin{tabular}{|c|c|c|c|c|c|c|}
\hline Bits & Message j & $X_{s_1}(j)$ & $\mathcal{H}_j$ & $X_r(j)$ & $p(j)$& $p_0(j)$ \\ 
\hline 00 & 1 & $s_1$ & $\lbrace2,4\rbrace$ & $s_1$ &  $4$ & $4$ \\ 
\hline 01 & 2 & $s_2$ & $\lbrace1,3\rbrace$ & $s_3$ &  $4$ & $4$ \\ 
\hline 10 & 3 & $s_3$ & $\lbrace2,4\rbrace$ & $s_4$ &  $4$ & $4$ \\ 
\hline 11 & 4 & $s_4$ & $\lbrace3,1\rbrace$ & $s_2$ &  $4$ & $4$ \\ 
\hline 
\end{tabular} 
\end{table*}
\begin{table*}
\centering
\caption{8-PSK, NODF SCHEME}
\label{table:table3}
\begin{tabular}{|c|c|c|c|c|c|c|c|c|}
\hline Bits & Message j & $X_{s_1}(j)$ & $\mathcal{H}_j$ & $X_r(j)$ & $\mathcal{K}_j$ & $X_{s_2}(j)$ & $p(j)$ & $p_0(j)$ \\ 
\hline 000 & 1 & $s_1$ & $\lbrace2,8\rbrace$ & $s_1$ & $\lbrace3\rbrace$ & $s_1$ & $1.9716$ & $0.3775$ \\ 
\hline 001 & 2 & $s_2$ & $\lbrace1,3\rbrace$ & $s_5$ & $\lbrace8\rbrace$ & $s_3$ & $1.9716$ & $0.3775$ \\ 
\hline 010 & 3 & $s_3$ & $\lbrace2,4\rbrace$ & $s_2$ & $\lbrace1,5\rbrace$ & $s_5$ & $1.8544$ & $0.3775$ \\ 
\hline 011 & 4 & $s_4$ & $\lbrace3,5\rbrace$ & $s_7$ & $\lbrace6\rbrace$ & $s_6$ & $1.9716$ & $0.3775$ \\ 
\hline 100 & 5 & $s_5$ & $\lbrace4,6\rbrace$ & $s_3$ & $\lbrace3,7\rbrace$ & $s_8$ & $1.8544$ & $0.3775$ \\ 
\hline 101 & 6 & $s_6$ & $\lbrace5,7\rbrace$ & $s_8$ & $\lbrace4\rbrace$ & $s_2$ & $1.9716$ & $0.3775$ \\ 
\hline 110 & 7 & $s_7$ & $\lbrace6,8\rbrace$ & $s_4$ & $\lbrace5,8\rbrace$ & $s_4$ & $1.3716$ & $0.3775$ \\ 
\hline 111 & 8 & $s_8$ & $\lbrace7,1\rbrace$ & $s_6$ & $\lbrace7,2\rbrace$ & $s_7$ & $1.3716$ & $0.3775$ \\ 
\hline 
\end{tabular} 
\end{table*}
\begin{table*}
\centering
\caption{8-PSK, ODF SCHEME}
\label{table:table4}
\begin{tabular}{|c|c|c|c|c|c|c|c|c|}
\hline Bits & Message j & $X_{s_1}(j)$ & $\mathcal{H}_j$ & $X_r(j)$ &  $p(j)$& $p_0(j)$ \\ 
\hline 000 & 1 & $s_1$ & $\lbrace2,8\rbrace$ & $s_1$   & $1.1716$ & $0.3431$ \\ 
\hline 001 & 2 & $s_2$ & $\lbrace1,3\rbrace$ & $s_5$   & $1.1716$ & $0.3431$ \\ 
\hline 010 & 3 & $s_3$ & $\lbrace2,4\rbrace$ & $s_2$   & $1.1716$ & $0.3431$ \\ 
\hline 011 & 4 & $s_4$ & $\lbrace3,5\rbrace$ & $s_7$   & $1.1716$ & $0.3431$ \\ 
\hline 100 & 5 & $s_5$ & $\lbrace4,6\rbrace$ & $s_3$   & $1.1716$ & $0.3431$ \\ 
\hline 101 & 6 & $s_6$ & $\lbrace5,7\rbrace$ & $s_8$   & $1.1716$ & $0.3431$ \\ 
\hline 110 & 7 & $s_7$ & $\lbrace6,8\rbrace$ & $s_4$   & $1.1716$ & $0.3431$ \\ 
\hline 111 & 8 & $s_8$ & $\lbrace7,1\rbrace$ & $s_6$   & $1.1716$ & $0.3431$ \\ 
\hline 
\end{tabular} 
\end{table*}
\begin{example}
\begin{figure}[htbp]
\centering
\includegraphics[totalheight=2in,width=4in]{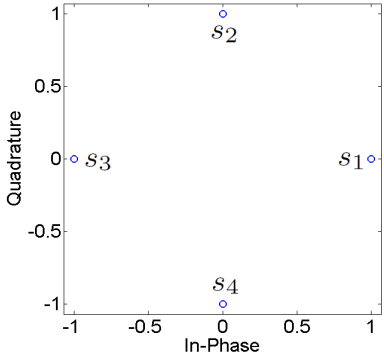}
\caption{4-PSK signal set}	
\label{fig:4psk}	
\end{figure}
We consider the case where 4-PSK is the signal set used at S and R whose points are labelled as shown in Fig. \ref{fig:4psk}. The value of $\alpha$ is assumed to be 0.1. The choice of the labelling scheme is described below.
\begin{itemize}
\item Choice of $X_{s_1}$ :
The map $X_{s_1}$ can be chosen arbitrarily. We can choose $X_{s_1}(i)=s_i$, $1 \leq i \leq 4$. 
\newline
The sets $\mathcal{H}_i$, $1 \leq i \leq 4$, can be found and are shown in Table \ref{table:table1}.
\item Choice of $X_{r}$ :
\begin{enumerate}
\item $X_r(1)$ can be chosen to be $s_1$.
\item The set $\mathcal H_1 = \lbrace{2,4}\rbrace$. We choose $X_r(2)=s_3$, since the Euclidean distance  between $s_3$ and $s_1$ is maximum. 
\item $X_r(4)$ can take only two possible symbols $s_2$ and $s_4$, both of which are at a squared Euclidean distance $\delta$ from $s_1$. Hence the value of $\min_{a,\bar{a}}m_1(a,\bar{a})m_2(a,\bar{a})$ cannot be made greater than $\delta^2$. $X_r(4)$ and  $X_r(3)$ are chosen to be $s_2$ and $s_4$ respectively.
\end{enumerate}
The sets $\mathcal{K}_i$, $1 \leq i \leq 4$, can be found and are shown in Table \ref{table:table1}.
\item Choice of $X_{s_2}$ :
\begin{enumerate}
\item $X_{s_2}(1)$ can be chosen to be $s_1$.
\item The set $\mathcal K_1 = \lbrace{4}\rbrace$. Hence choose $X_{s_2}(4)=s_3$.
\item The set $\mathcal K_4 = \lbrace{1}\rbrace$, for which symbol has already been assigned.
\item Since the sets $\mathcal K_1$ and $\mathcal K_4$, both do not contain $2$, the choice of $X_{s_2}(2)$ can be made arbitrarily. We choose $X_{s_2}(2)=s_2$. As a result $X_{s_2}(3)=s_4$.
\end{enumerate}
\end{itemize}

The choice of $X_{s_1}$, $X_{r}$ and $X_{s_2}$ thus made is tabulated in Table \ref{table:table1}. Table \ref{table:table1} also contains $p(a)$ and $p_0(a)$ (defined in the beginning of this section), for $1 \leq a \leq 4$. For the labelling scheme $\mathcal{L}_0$, the maps $X_r$ and $X_{s_2}$ are taken to be same as $X_{s_1}$.
From Table \ref{table:table1}, we see that for the NODF scheme $d(\mathcal{L})= \min_a p(a)=4.8$ and $d(\mathcal{L}_0)= \min_a p_0(a)=4.4$.
Hence the labelling gain,  $L_G^{NODF}(\mathcal{L})=\dfrac{4.8}{4.4}= 0.3779$ dB.

Similarly, from Table \ref{table:table2}, we see that for the ODF scheme $d(\mathcal{L})= \min_a p(a)=4$ and $d(\mathcal{L}_0)= \min_a p_0(a)=4$.
Hence the labelling gain, $L_G^{ODF}(\mathcal{L})=1=0$ dB.
\end{example}

\begin{example}
\begin{figure}[htbp]
\centering
\includegraphics[totalheight=2in,width=4in]{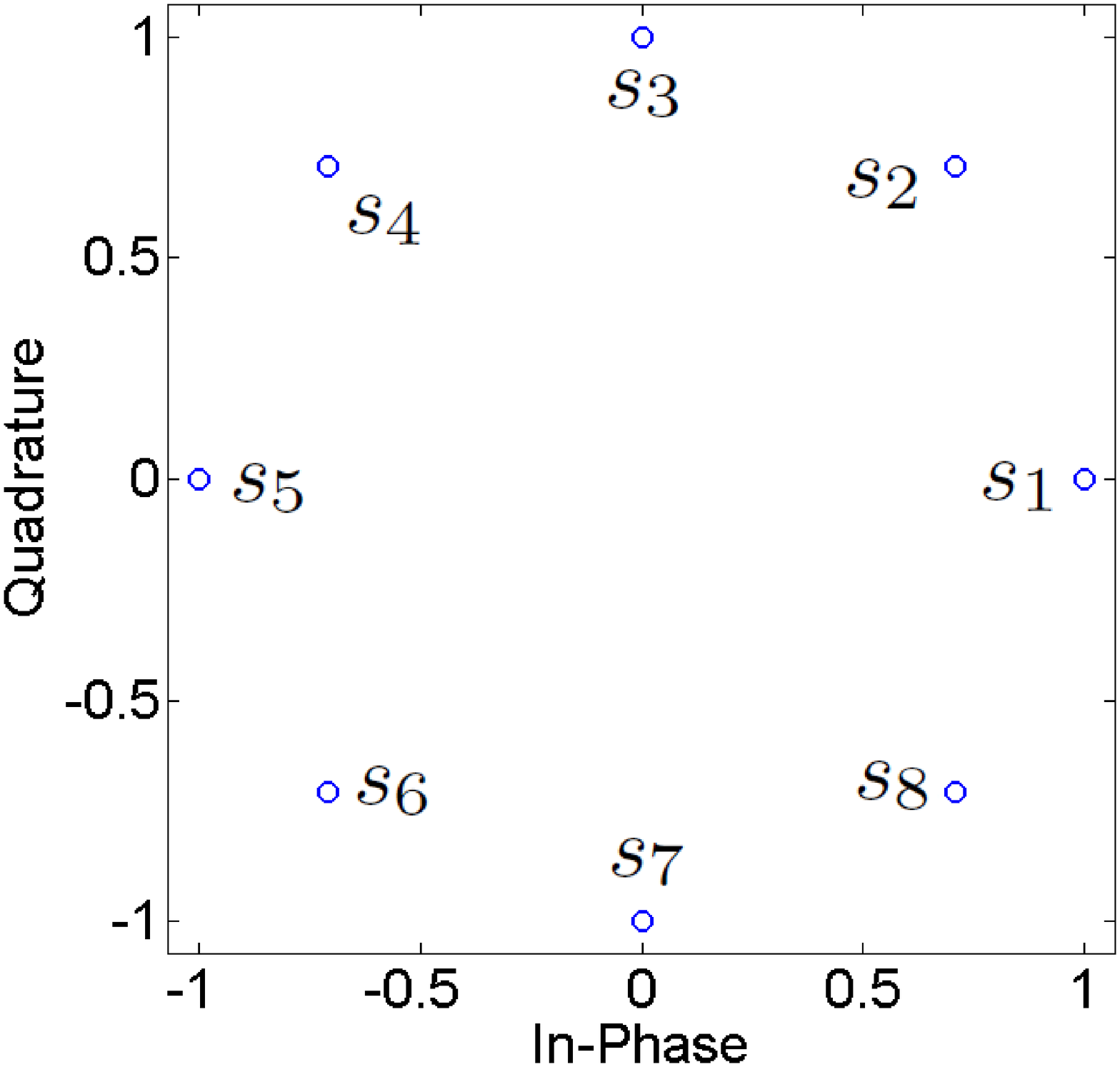}
\caption{8-PSK signal set}	
\label{fig:8psk}	
\end{figure}
Consider the case where 8-PSK is the signal set used at S and R. The points are assigned labels as shown in Fig. \ref{fig:8psk}. The value of $\alpha$ is taken to be 0.1.
\begin{itemize}
\item Choice of $X_{s_1}$ :
The map $X_{s_1}$ can be chosen arbitrarily. We can choose $X_{s_1}(i)=s_i$, $1 \leq i \leq 8$. 
\newline
The sets $\mathcal{H}_i$, $1 \leq i \leq 8$, can be found and are shown in Table \ref{table:table3}.
\item Choice of $X_{r}$ :
\begin{enumerate}
\item $X_r(1)$ can be chosen to be $s_1$.
\item The set $\mathcal H_1 = \lbrace{2,8}\rbrace$. We choose $X_r(2)=s_5$, since the Euclidean distance  between $s_5$ and $s_1$ is maximum. We choose $X_r(8)=s_6$, since the Euclidean distance  between $s_6$ and $s_1$ is maximum, among all possible symbols which are not assigned. 
\item The set $\mathcal H_2 = \lbrace{1,3}\rbrace$. $X_r(3)$ is chosen to be $s_2$, since its Euclidean distance  from $s_5$ is maximum, among all symbols which are not yet assigned.
\item 
The set $\mathcal H_3 = \lbrace{2,4}\rbrace$ and hence $X_r(4)$ is chosen to be $s_3$.
\item
$\mathcal H_4 = \lbrace{3,5}\rbrace$ and hence $X_r(5)$ is chosen to be $s_7$.
\item
$\mathcal H_5 = \lbrace{4,6}\rbrace$ and hence $X_r(6)$ is chosen to be $s_8$. Finally we are left with $X_r(7)=s_4$. 
\end{enumerate}
Since the steps described above results in a value of $\min_{a,\bar{a}}m_1(a,\bar{a})m_2(a,\bar{a}) \gneq \delta^2$, the process of assigning the map $X_r$ is complete. The sets $\mathcal{K}_i$, $1 \leq i \leq 8$, can be found and are shown in Table \ref{table:table3}.
\item Choice of $X_{s_2}$ :
\begin{enumerate}
\item $X_{s_2}(1)$ can be chosen to be $s_1$.
\item The set $\mathcal K_1 = \lbrace{3}\rbrace$. Hence choose $X_{s_2}(3)=s_5$.
\item The set $\mathcal K_3 = \lbrace{1,5}\rbrace$. Hence choose $X_{s_2}(5)=s_8$, since its Euclidean distance from $s_5$ is maximum.
\item The set $\mathcal K_5 = \lbrace{3,7}\rbrace$. Hence choose $X_{s_2}(7)=s_4$.
\item The set $\mathcal K_7 = \lbrace{5,8}\rbrace$. Hence choose $X_{s_2}(8)=s_7$.
\item The set $\mathcal K_8 = \lbrace{5,8}\rbrace$. Hence choose $X_{s_2}(2)=s_3$.
\item The set $\mathcal K_2 = \lbrace{8}\rbrace$, for which symbol has been already assigned.
\item We are left with messages $4$ and $6$. $\mathcal K_4 = \lbrace{6}\rbrace$ and $\mathcal K_6 = \lbrace{4}\rbrace$. Choose $X_{s_2}(4)=s_6$ and $X_{s_2}(6)=s_2$.
\end{enumerate}
\end{itemize}
\begin{figure}[htbp]
\centering
\includegraphics[totalheight=2.5in,width=3.75in]{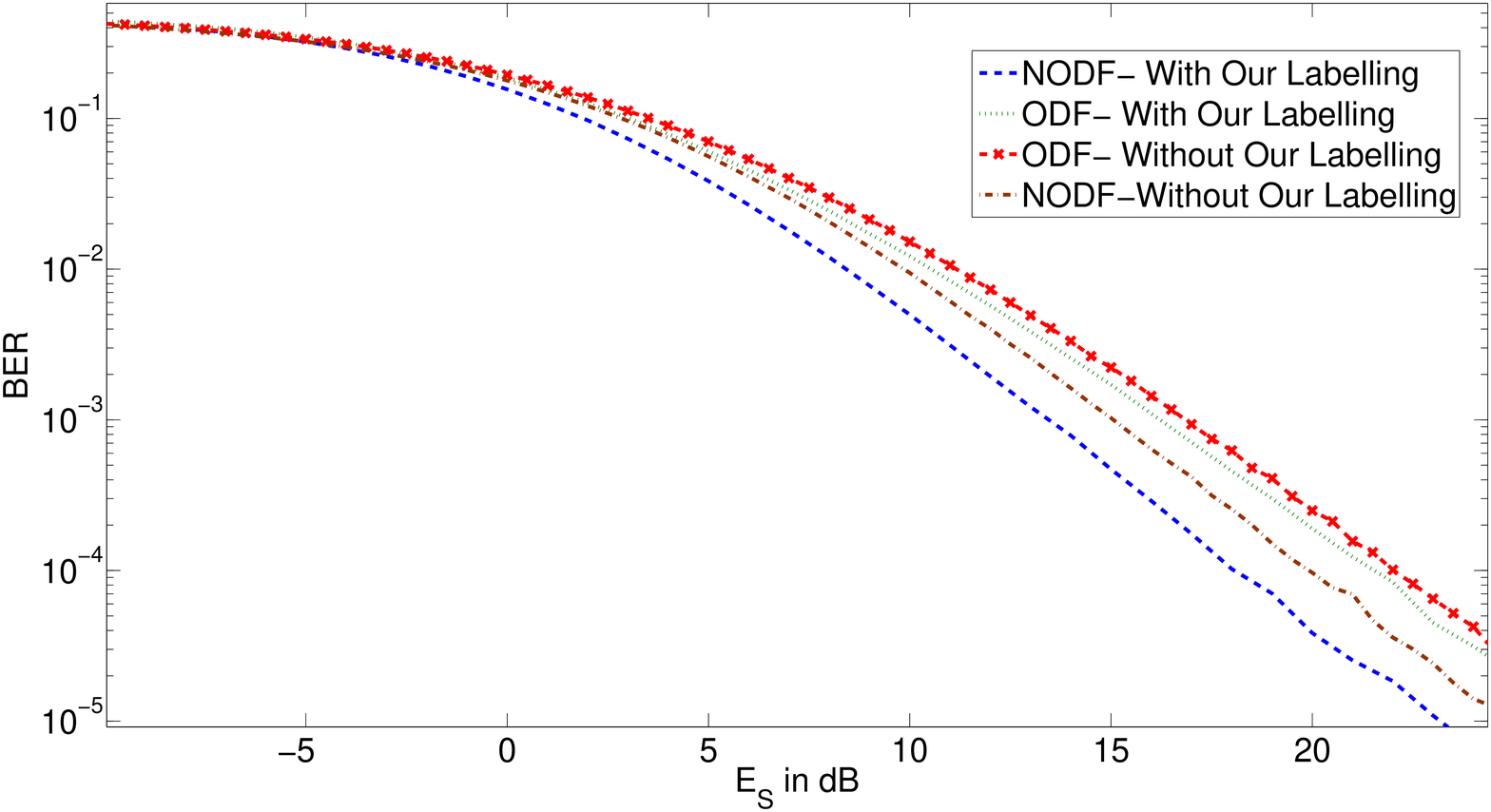}
\caption{$E_{\mathcal{S}}$ Vs $BER$ performance of NODF and ODF schemes, with and without our labelling for 8-PSK, with $\sigma_{ds}=0$ dB, $\sigma_{rs}=10$ dB and $\sigma_{dr}=10$ dB.}	
\label{fig:fig3}	
\end{figure}
The choice of $X_{s_1}$, $X_{r}$ and $X_{s_2}$ thus made is tabulated in Table \ref{table:table3}.
From Table \ref{table:table3}, we see that for the NODF scheme $d(\mathcal{L})= \min_a p(a)=1.3716$ and $d(\mathcal{L}_0)= \min_a p_0(a)=0.3775$.
Hence the labelling gain, 
\begin{align}
\centering
\nonumber
L_G^{NODF}(\mathcal{L})=\dfrac{1.3716}{0.3775}= 5.6031 dB.
\end{align}
Similarly, from Table \ref{table:table2}, we see that for the ODF scheme $d(\mathcal{L})= \min_a p(a)=1.1716$ and $d(\mathcal{L}_0)= \min_a p_0(a)=0.3431$.
Hence the labelling gain,
\begin{align}
\centering
\nonumber
L_G^{ODF}(\mathcal{L})=\dfrac{1.1716}{0.3431}= 5.3336 dB.
\end{align}

Simulation results showing the $E_{\mathcal{S}}$ Vs $BER$ performance of the NODF and ODF schemes, with our labelling and without our labelling, with 8-PSK as the constellation used at S and R, is shown in Fig. \ref{fig:fig3}. From Fig. \ref{fig:fig3}, it can be seen that for both the ODF and the NODF schemes, the labelling strategy suggested in this section provides advantage. For the ODF scheme, at high SNR, the gain provided by the labelling scheme described is about 0.5 dB and for the NODF scheme, it is about 2 dB. Consistent with the observations made in the beginning of this section based on the PEP expressions for the ODF and NODF schemes, from Fig. \ref{fig:fig3}, it can be seen that the gain provided by the choice of the labelling is more in the case of the NODF scheme than the ODF scheme.
\end{example}
\section{COMPARISON WITH RELAY CHANNEL WITH IDEAL S-R LINK}
\begin{figure*}
\scriptsize
\begin{align}
\label{eqn15}
P_a(a \longrightarrow \bar{a}\vert c_{rs},c_{ds_1},c_{ds_2})&=Q\left[\dfrac{\sqrt{\vert c_{ds_1}\left(X_{s_1}\left(a\right)-X_{s_1}\left(\bar{a}\right)\right)\vert^2+ \vert c_{ds_2}\left(X_{s_2}(a)-X_{s_2}\left(\bar{a}\right)\right)+c_{dr}\left(X_{r}(a)-X_{r}(\bar{a})\right)\vert^2}}{\sqrt{2}}\right]\\
\label{eqn16}
&\leq \exp \left\lbrace -\dfrac{\vert c_{ds_1}\left(X_{s_1}\left(a\right)-X_{s_1}\left(\bar{a}\right)\right)\vert^2+ \vert c_{ds_2}\left(X_{s_2}(a)-X_{s_2}\left(\bar{a}\right)\right)+c_{dr}\left(X_{r}(a)-X_{r}(\bar{a})\right)\vert^2}{4}\right\rbrace\\
\label{eqn17}
&\hspace{-3 cm}Pr \left(a \longrightarrow \bar{a} \right) \leq 
\hspace{0 cm} \left [\dfrac{1}{1+\dfrac{1}{4}\vert \sigma_{ds}\vert ^2\vert X_{s_1}({a})-X_{s_1}(\bar{a})\vert ^2}\right]
\left[\dfrac{1}{1+\dfrac{1}{4}\vert \sigma_{ds}\vert ^2\vert X_{s_2}({a})-X_{s_2}(\bar{a})\vert ^2+\dfrac{1}{4}\vert \sigma_{dr}\vert ^2\vert X_{r}({a})-X_{r}(\bar{a})\vert ^2}\right]
\end{align} 
\hrule
\end{figure*}
We consider the case where the S-R relay link is ideal, i.e it is assumed that R decodes the message it receives with zero probability of error. The optimal ML decoder for this case is
\begin{align}
\nonumber
\hat{a}_{ML}&=\arg\max_a Pr\left(y_{d_1},y_{d_2}\vert m=a, c_{ds_1}, c_{ds_2}, c_{dr}\right) \\
\nonumber
&=\arg\min_a (\vert y_{d_1}-c_{ds_1}X_{s_1}(a)\vert^2 \\
\nonumber
&\hspace{2 cm}+ \vert y_{d_2}-c_{ds_2}X_{s_2}(a)-c_{dr}X_{r}(a)\vert^2).
\end{align}
\begin{figure}[htbp]
\centering
\includegraphics[totalheight=2.5in,width=3.75in]{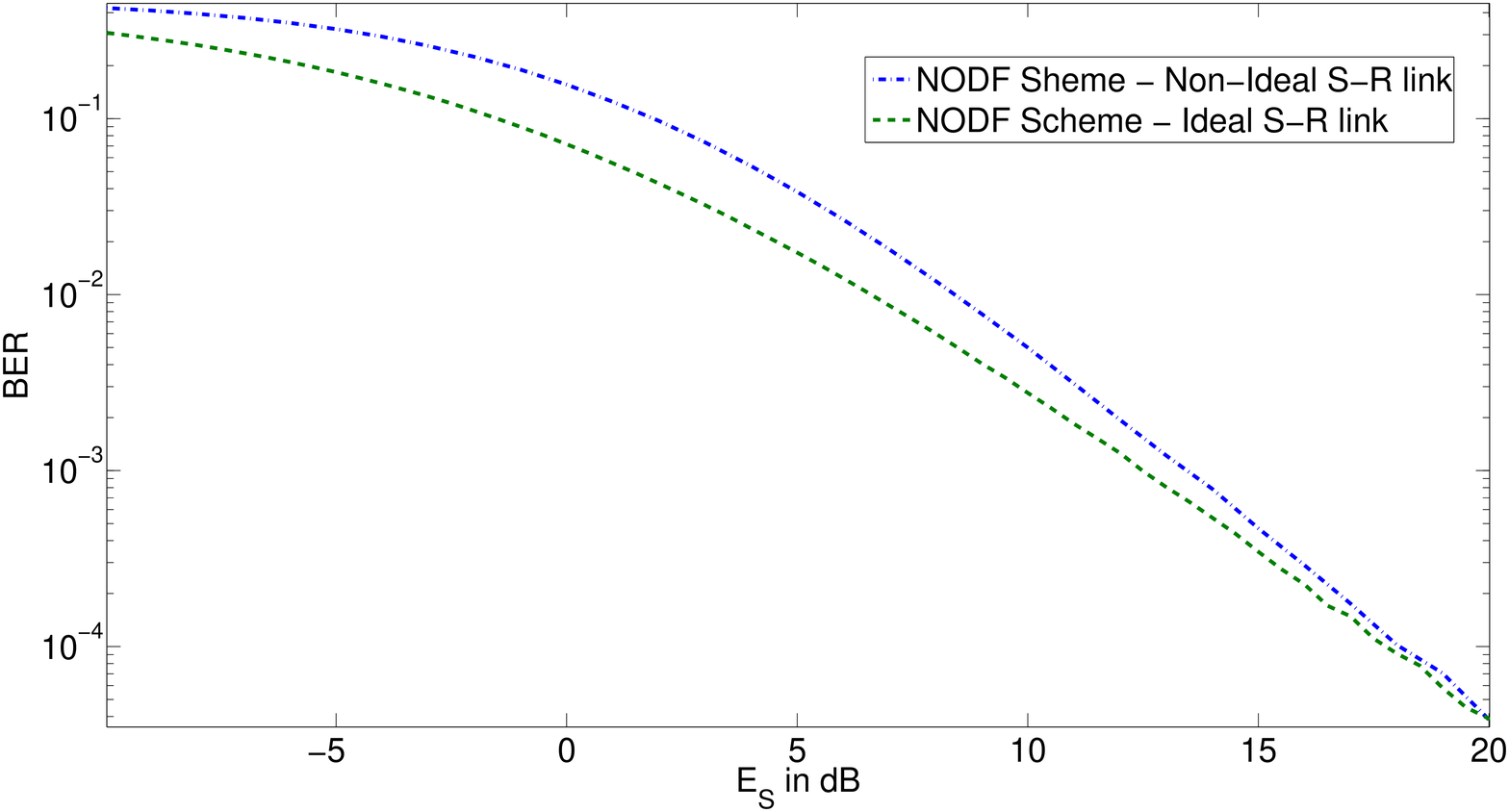}
\caption{$E_{\mathcal{S}}$ Vs $BER$ performance of the NODF scheme with our labelling, with ideal and non-ideal S-R links for 8-PSK}	
\label{fig:fig4}	
\end{figure}
\begin{figure}[htbp]
\centering
\includegraphics[totalheight=2.5in,width=3.75in]{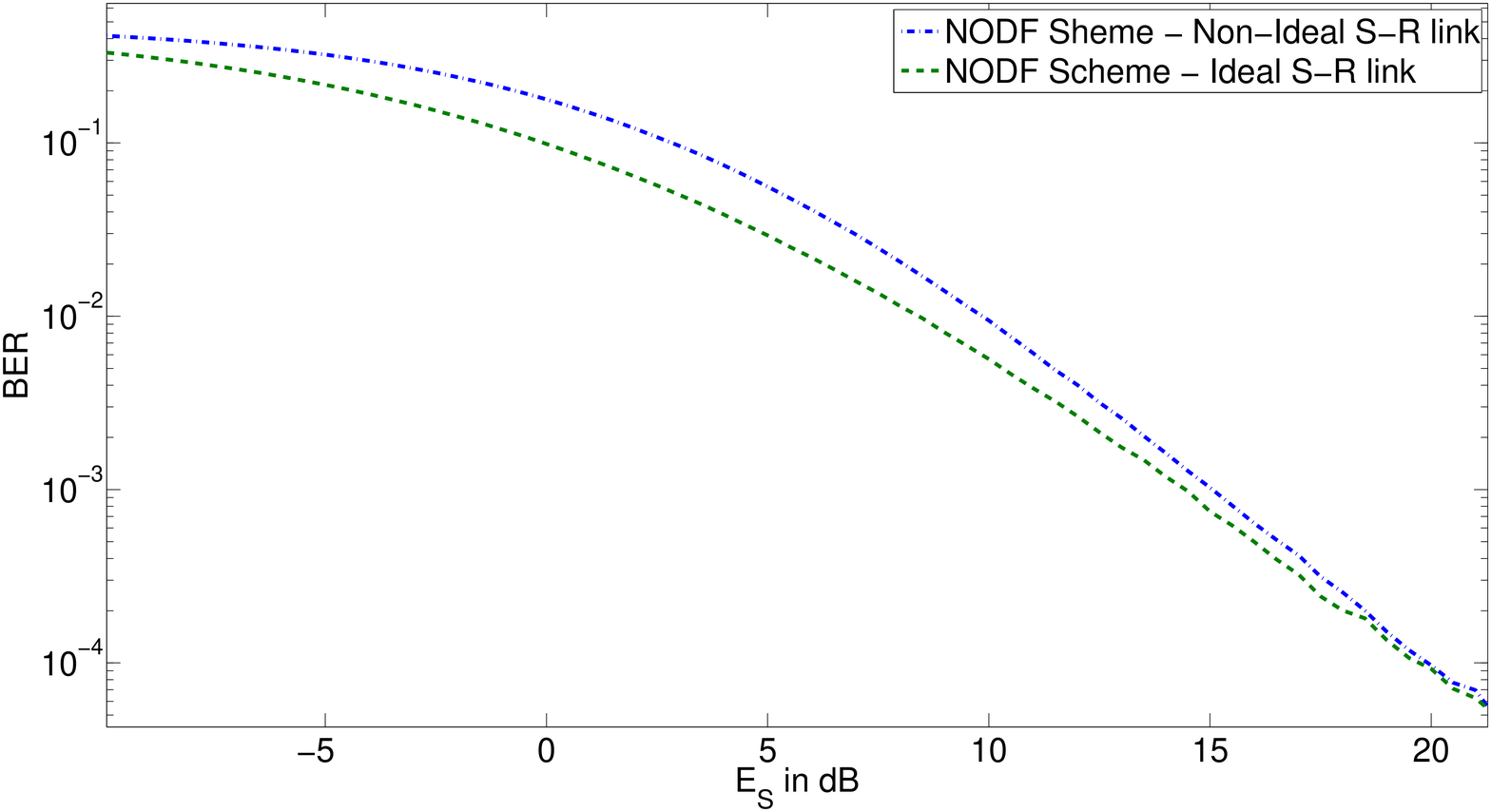}
\caption{$E_{\mathcal{S}}$ Vs $BER$ performance of the NODF scheme without our labelling, with ideal and non-ideal S-R links for 8-PSK}	
\label{fig:fig5}	
\end{figure}
\begin{figure}[htbp]
\centering
\includegraphics[totalheight=2.5in,width=3.75in]{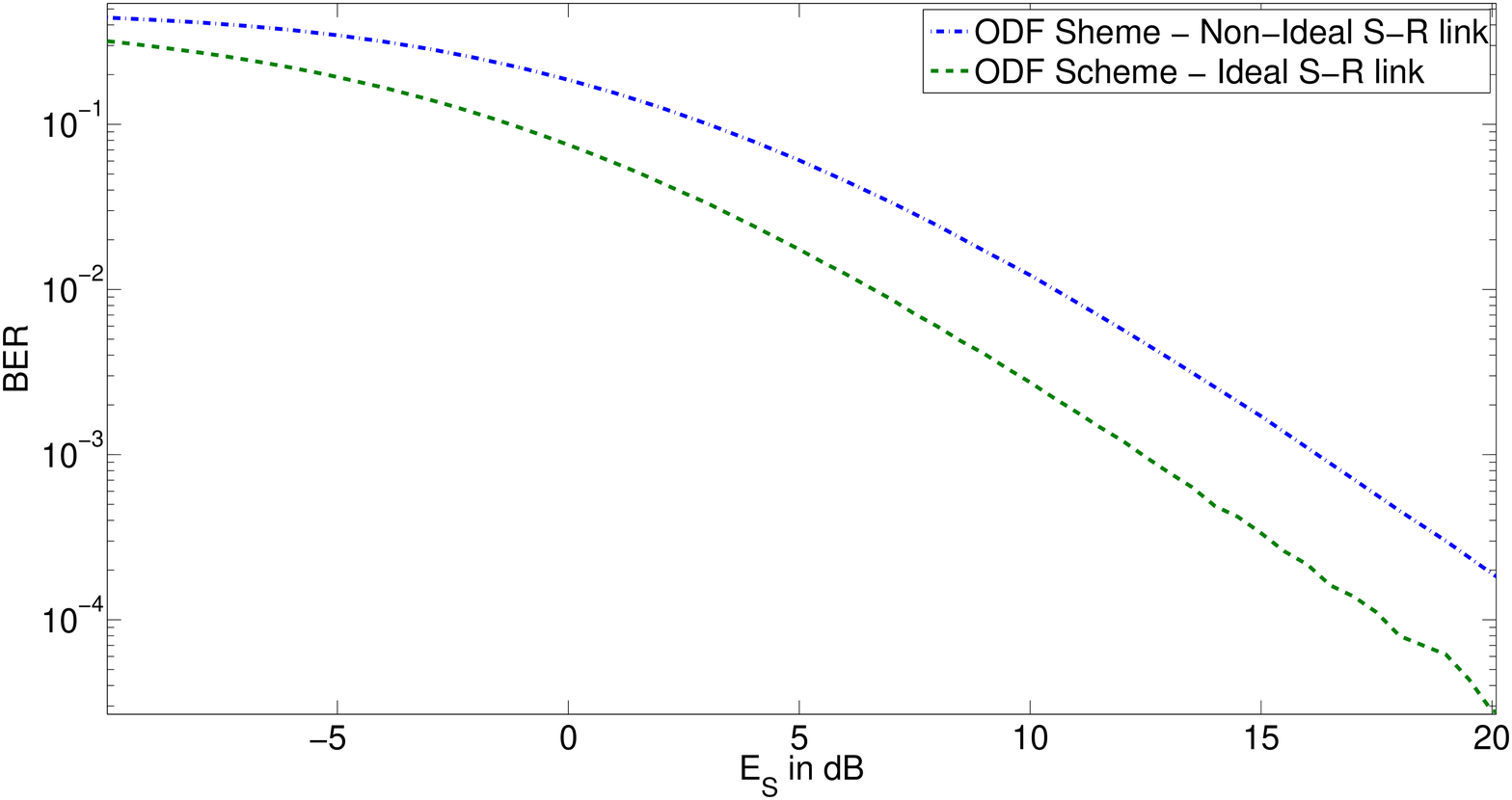}
\caption{$E_{\mathcal{S}}$ Vs $BER$ performance of the ODF scheme with our labelling, with ideal and non-ideal S-R links for 8-PSK}	
\label{fig:fig6}	
\end{figure}
\begin{figure}[htbp]
\centering
\includegraphics[totalheight=2.5in,width=3.75in]{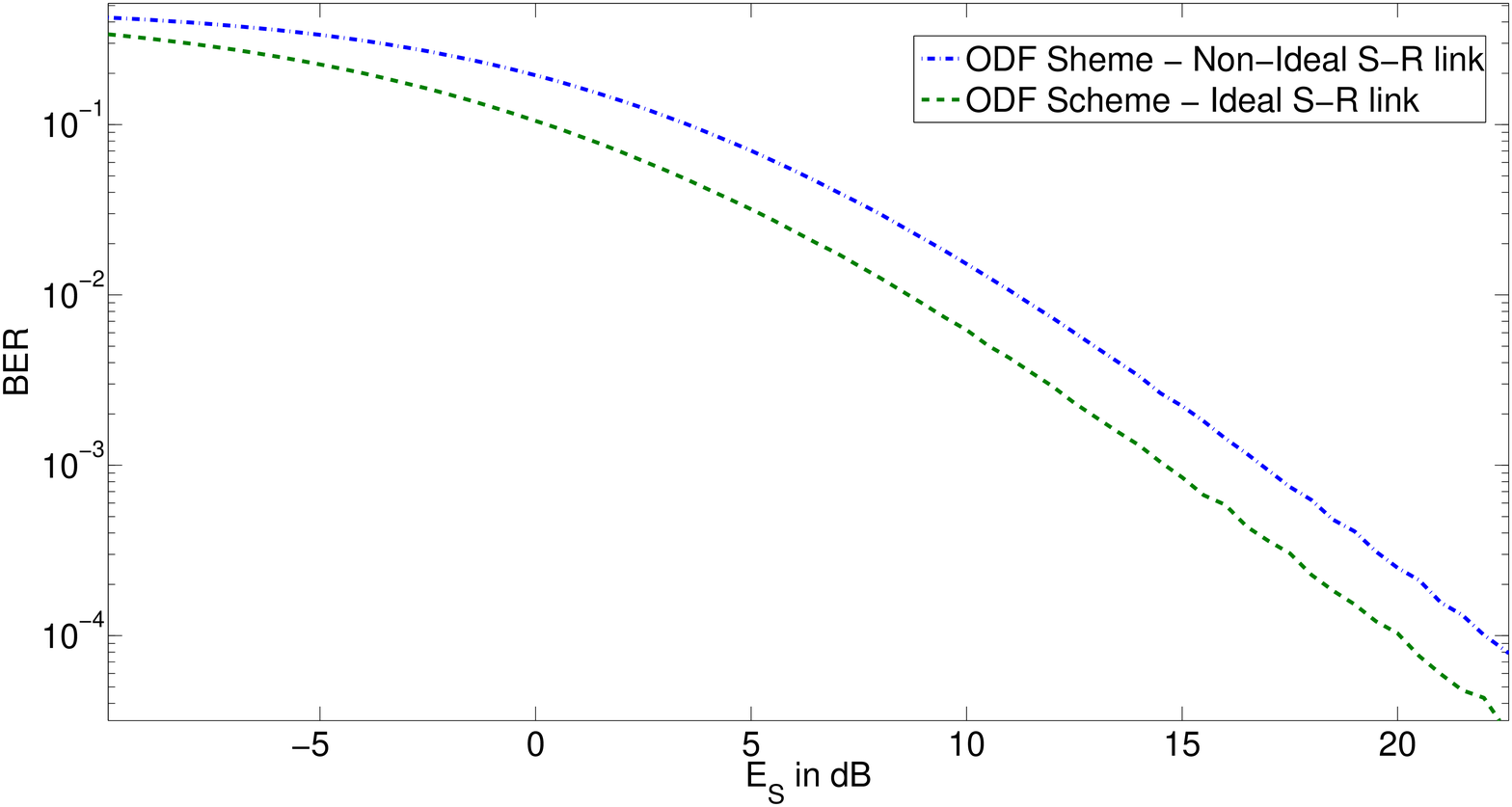}
\caption{$E_{\mathcal{S}}$ Vs $BER$ performance of the NODF scheme without our labelling, with ideal and non-ideal S-R links for 8-PSK}	
\label{fig:fig7}	
\end{figure}
The PEP that message $a$ transmitted by S is decoded as message $\bar{a}$ by D is given by \eqref{eqn15}.
Taking expectation of \eqref{eqn16} with respect to $c_{ds_1}$, $c_{ds_2}$ and $c_{dr}$, we get \eqref{eqn17}.
We note that at high SNR, the bound on the PEP given by \eqref{eqn17} and Theorem 1 (neglecting the higher order terms) are the same. Hence at high SNR, the performance of the NODF scheme with a non-ideal S-R link is expected to be same as that of the NODF scheme with an ideal S-R link. In others words, at high SNR, the $E_{\mathcal{S}}$ Vs $BER$ performance does not depend on the strength of the S-R link. On the other hand, the PEP bound for the ODF scheme given in Corollary 1 contains additional second order terms and is not the same as the one obtained by substituting $X_{s_1}=X_{s}$ and  $X_{s_2}=0$ in \eqref{eqn17}. Hence at high SNR, the $E_{\mathcal{S}}$ Vs $BER$ performance of the ODF scheme with a non-ideal S-R link is not expected to be be the same as that of the ODF scheme with an ideal S-R link.  

A comparison of $E_{\mathcal{S}}$ Vs $BER$ performance of the NODF scheme with our labelling, for the cases where S-R link is ideal and non-ideal is shown in Fig. \ref{fig:fig4}. A similar comparison for the NODF scheme without our labelling is presented in Fig. \ref{fig:fig5}. From Fig. \ref{fig:fig4} and Fig \ref{fig:fig5}, it is seen clearly that at high SNR the performance of the NODF schemes with a non-ideal S-R link and ideal S-R link exactly coincide.
In Fig. \ref{fig:fig6} and Fig \ref{fig:fig7}, similar comparisons are made for the ODF scheme with and without our labelling. From,  Fig. \ref{fig:fig6} and Fig \ref{fig:fig7}, it can be seen that at high SNR, the $E_{\mathcal{S}}$ Vs $BER$ curves for the case where the S-R link is ideal and non-ideal do not coincide, unlike the NODF scheme. In other words, to study the high SNR performance of the NODF scheme, we can assume the S-R link to be ideal, whereas the same is not true for the  ODF scheme.

\section{DISCUSSION} 
A near ML decoder which gives maximum possible diversity (diversity order 2) was studied. It was shown that the NODF scheme provides advantage over the ODF scheme. A proper choice of the labelling scheme used at the source and the relay results in a significant improvement in performance. It will be interesting to study the performance of the near ML decoder and the effect of the choice of labelling, when the source and the relay use coded communication techniques. 

\section*{Acknowledgement}
 This work was supported  partly by the DRDO-IISc program on Advanced Research in Mathematical Engineering through a research grant as well as the INAE Chair Professorship grant to B.~S.~Rajan.

\end{document}